\documentclass[aps,pra,twocolumn,floatfix,showpacs]{revtex4-1}

\usepackage{graphicx}
\usepackage{dcolumn}
\usepackage{bm}     
\usepackage[margin=2.0cm]{geometry}
\usepackage{amsmath}
\usepackage{braket}
\usepackage{appendix}

\begin{document}
	\title{Factorization of high-harmonic generation yields in impurity-doped materials}
	\author{Van-Hung Hoang$^{1,2,3}$, Anh-Thu Le$^{1}$}
	\affiliation{$^1$Department of Physics, Missouri University of Science and Technology, Rolla, MO 65409, USA}
	\affiliation{$^2$Atomic Molecular and Optical Physics Research Group, Advanced Institute of Materials Science, Ton Duc Thang University, Ho Chi Minh City, Vietnam}
	\affiliation{$^3$Faculty of Applied Sciences, Ton Duc Thang University, Ho Chi Minh City, Vietnam}
	\email{hoangvanhung@tdtu.edu.vn}
	\begin{abstract}
		We present a theoretical investigation of high-harmonic generation (HHG) from impurity-doped materials using the time-dependent Schr\"odinger equation (TDSE) approach. We demonstrate the factorization of HHG yields as a product of an electron wave packet and the recombination cross section, in analogy to HHG from atoms and molecules in the gas phase. Furthermore, we show that the quantitative rescattering model based on this factorization accurately reproduces the TDSE results. This opens up new possibilities to study impurities in materials using the available techniques from strong-field physics. 
	\end{abstract}

\maketitle

High-harmonic generation (HHG) from atoms and molecules in gas phase is one of the most important phenomena in intense laser-matter interactions. It provides not only table-top coherent XUV to soft X-ray light sources, but also a powerful technique to produce attosecond pulses for applications in science and technology \cite{Krausz:rmp09}. It has been studied over the last three decades and the mechanism behind HHG in gases is now well understood based on the three-step model \cite{Corkum:prl93,Krause:prl92}. Only recently, HHG in solids was demonstrated experimentally with mid-infrared \cite{Ghimire:NatPhys10,Vampa:Nature15}, visible \cite{Luu:Nature15}, and terahertz lasers \cite{Zaks:Nature12,Schubert:NatPhot14,Hohenleutner:Nature15}. Subsequently, HHG has been observed in variety of materials, including wide bandgap dielectrics \cite{Ndabashimiye:Nature16,You:NatPhys17,Luu:NatCom18}, amorphous common glass \cite{You:NatCom17}, and graphene \cite{Yoshikawa:Science17}. Due to their high atomic density, it has been expected that solids have the potential to produce more efficient HHG, as compared to atoms and molecules in the gas phase. Quite recently, it was demonstrated both experimentally and theoretically that HHG from solids can be enhanced if the material is doped by impurities \cite{Sivis:Science17,Huang:pra17,Almalki:prb18,Madsen:pra19}. In fact, impurities or target engineering in general are typically used to alter the band structures of a solid, therefore allowing one to actively control various processes in the material. Conceptually, it was shown that the three-step recollision model can be extended for HHG from impurities in solids \cite{Almalki:prb18,Madsen:pra19}. This indicates that HHG process from impurities in solids is very similar to that from gases. 

In this paper we show that HHG yields from impurities can be expressed as a product of a returning electron wave-packet and the photo-recombination cross section for electron in conduction bands back to the impurity ground state. This indicates that the quantitative rescattering theory (QRS) for atoms and molecules in the gas phase \cite{Morishita:prl08,Le:pra09,Lin:jpb10,Le:jpb16,Frolov:prl09,Frolov:jpb09} can be extended to HHG from impurities in solids. Our results therefore confirm the validity of the three-step model and recollision picture for HHG from impurities \cite{Almalki:prb18,Madsen:pra19} and elevate it to a more quantitative level. 
                           
To simulate a HHG process, we solve the time-dependent Schr\"odinger equation (TDSE) for solids in intense laser pulse within the single-active-electron approximation. This approach has been used by different groups \cite{Vampa:prl14,Hawkins:pra15,Higuchi:prl14,Wu:pra15,Guan:pra16,Du:pra18,Navarrete:pra19}. Electron correlation has typically been taken into account empirically via relaxation times in the semiconductor Bloch equations approach \cite{Vampa:prl14,Luu:prb16,Jiang:prl18,Floss:pra18}. Note that an approach based on the solution of the time-dependent density functional theory (TDDFT) has also been used, in which the electron exchange and correlation can be taken into account \cite{Rubio:prl17,Bauer:prl18,Floss:pra18,Madsen:pra19,Yu:pra19}. In particular, it was shown that the use of the frozen Kohn-Sham potential leads to quite similar results as of the full TDDFT calculations. An attempt to treat electron correlation was also reported for HHG in a strongly correlated system \cite{Silva:NatPhot18}.

\begin{figure}[t]
	\centering
	\fbox{\includegraphics[width=0.9\linewidth]{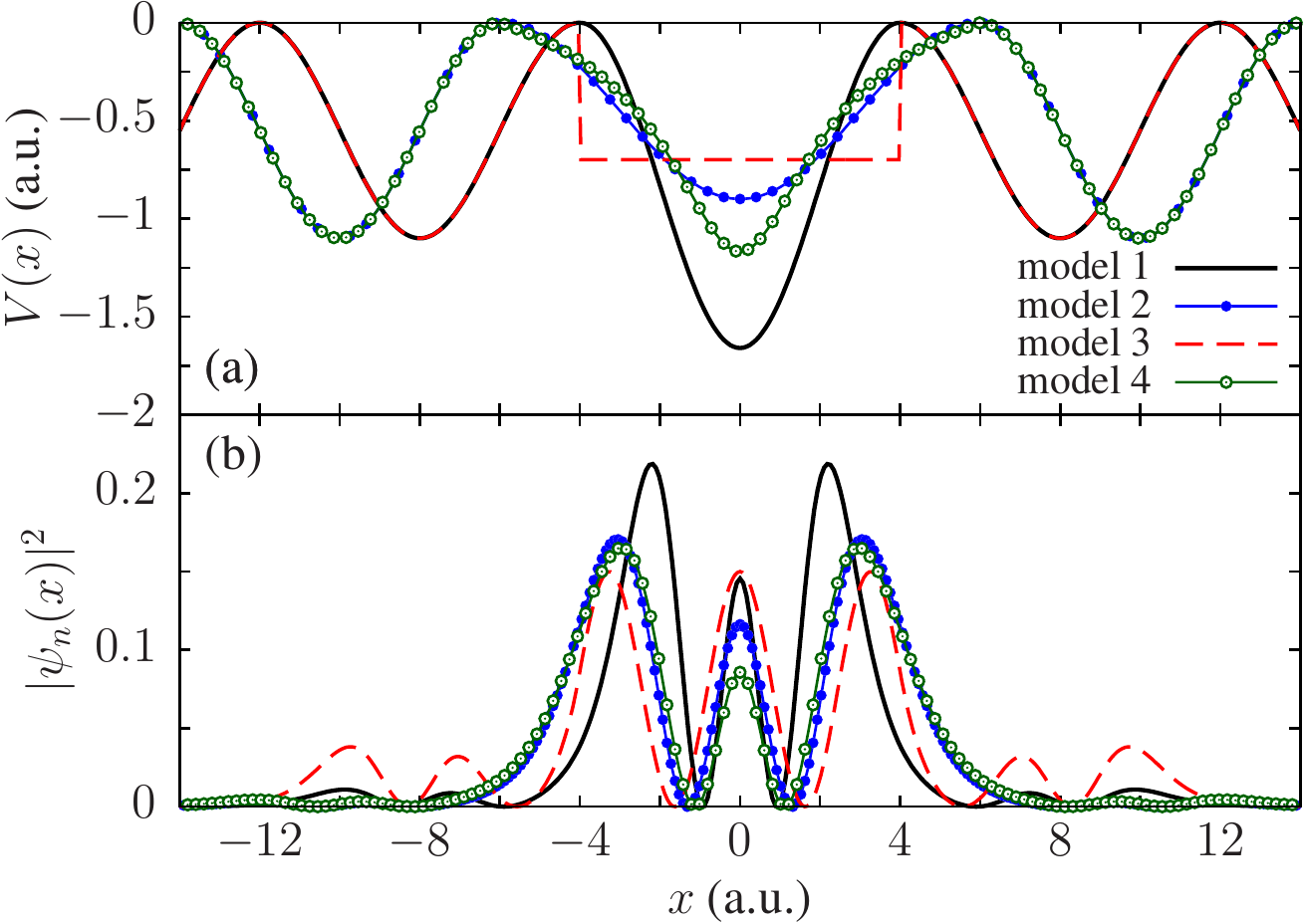}}
	\caption{The model potentials used in this paper (a) and the probability density of the impurity ground state orbitals for four models. The results are shown near the doping site, chosen to be at $x=0$.}
	\label{Fig1}
\end{figure}

We model the undoped solid as a linear chain of $N$ atoms located with a separation $a_0$. The effective potential for the active electron inside the undoped solid is modeled by a Mathieu-type potential as (atomic units are used throughout this paper, unless otherwise indicated)  
\begin{equation}
v(x) =  -v_0 \left[ 1+\cos(2\pi x/a_0) \right]. 
\end{equation}
In this paper we choose $N=101$, $v_0=0.45$, and the lattice constant $a_0=8$.
For the doped materials, we consider the case when dopant atoms substitute atoms of the undoped solid. To describe the effect of different impurity species on HHG spectra, we use four different model potentials for $v(x)$ near the doping site. We have limited ourselves to simple model potentials in order to illustrate the main physics. To be specific, we choose the doping site to be at $x=0$. The model potential is assumed to be modified only in the region $|x|\le a_1/2$. In this region, we use Mathieu-typle potential $v(x) =  -v_1 \left[ 1+\cos(2\pi x/a_1)\right]$ with parameters $v_1=0.83$ and $a_1=8$ and $v_1=0.45$ and $a_1=12$ for model 1 and 2, respectively. For model 3, we use $v(x) = - v_1$ with $v_1=0.7$ and $a_1=8$, and for model 4
\begin{equation}
v(x) = -\dfrac{v_1}{\sqrt{x^2+3}}  + \dfrac{v_1}{\sqrt{(a_1/2)^2+3}}, 
\end{equation}
with $v_1=2.8$ and $a_1=12$. The four model potentials near the impurity are shown in Fig.~\ref{Fig1}(a).  

\begin{figure}[htbp]
	\centering
	\fbox{\includegraphics[width=1.0\linewidth]{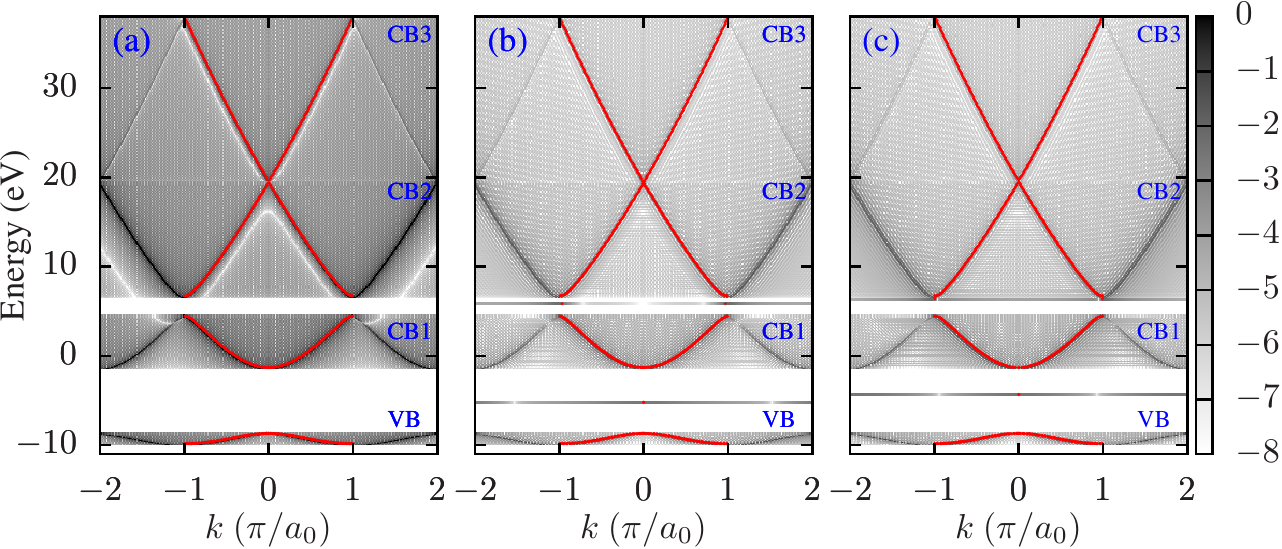}}
	\caption{Band structures of undoped solid (a) and doped solids with model 1 (b), and model 2 (c), as given by the probability densities for the eigenstates of $\hat{H}_0$ in the momentum space. The maximum of the probability densities for each energy is showed by a red dot. In (b) and (c), the isolated horizontal lines correspond to the impurity orbitals. The bands below the top valence band are not shown, since they practically are not involved in HHG process.}
	\label{Fig2}
\end{figure}

The time-independent Schr\"odinger equation can be written as 
\begin{equation}
\hat{H}_0\psi_n(x) = E_n\psi_n(x), 
\end{equation}
where $\hat{H}_0=\hat{p}^2/2+v(x)$. The above equation is solved by the discrete variable representation method with a uniform grid in the basis of Fourier functions \cite{Colbert:jcp92}. The probability density of the impurity ground state for the four models are showed in Fig.~\ref{Fig1}(b). Clearly, the impurity wave functions are mostly localized around the impurity site. Following Refs.~\cite{Hansen:pra17,Yu:pra19}, we calculate the band structures using the spatial Fourier transforms of the eigenfunctions of Hamiltonian $H_0$. This is illustrated in Fig.~\ref{Fig2}(a,b,c) for the undoped and doped solids with model 1 and model 2, respectively, where the probability densities in the momentum space ($k$-space) are plotted at their respective energies on a logarithmic scale. To verify the results, we also calculated the band structures using the Bloch state basis. The two results are identical. At such a low doping rate ($1\%$), the band structures do not change much and the impurity states are energetically isolated. This is consistent with the earlier results ~\cite{Yu:pra19}, in which the density functional theory was used. In all four cases, the impurity ground state energies are located around $-5$ eV, i.e., right in between the valence band (VB) and the first conduction band (CB1). Since the energy gap between the highest occupied orbital (the impurity ground state) and CB1 is much smaller than the band gap between the VB and CB1, it is expected that the HHG yields from doped solids in all four cases are significantly stronger than that from the undoped material. Our calculations based on the solutions of the TDSE indeed confirmed this expectation, in general agreements with results based on the TDDFT for donor-doped materials \cite{Yu:pra19}.

The time-dependent Hamiltonian for the doped material interacting with an intense laser pulse polarized along $x$-axis can be written in the length gauge within the dipole approximation as
\begin{equation}
\label{eqTDH}
\hat{H} = \hat{H}_0 +xE(t), 
\end{equation}
where the laser pulse is given by
\begin{equation}
E(t) = E_0 \cos^2\left(\dfrac{\pi t}{\tau}\right) \cos(\omega_0 t+\varphi) 
\end{equation}
for the time interval $(-\tau/2,\tau/2)$ and zero elsewhere. Here $E_0$ is the laser peak electric field amplitude, $\omega_0$ is the carrier frequency, and $\varphi$ is the carrier-envelope phase. For this choice of laser envelope, the pulse duration defined as the full width at half maximum (FWHM) of the intensity, is given as $\Gamma=\tau/2.75$. 

The TDSE with  Hamiltonian~\eqref{eqTDH} is solved by the split-operator method with the impurity ground-state wave function taken as the initial wave function. To avoid the unphysical reflection due to a finite box size, we use an absorbing boundary of the form of $\cos^{1/4}$. The numerical calculations are performed on a uniform spatial grid with $\Delta x=0.1$ and a time step $\Delta t=0.3$. We have checked these parameters to make sure that converged results were obtained.  
Once the time-dependent wave function is obtained, the time-dependent laser-induced currents $j(t)$ can be calculated. The HHG spectrum is then given as the modulus square of the Fourier-transformed of time-dependent laser-induced currents, where a window function $\cos^8\left(\dfrac{\pi t}{\tau}\right)$ has been applied before the Fourier transform is carried out \cite{Hansen:pra18}. We have also found that identical spectra are obtained by using the laser-induced dipoles. 

\begin{figure}[t]
	\centering
	\fbox{\includegraphics[width=0.9\linewidth]{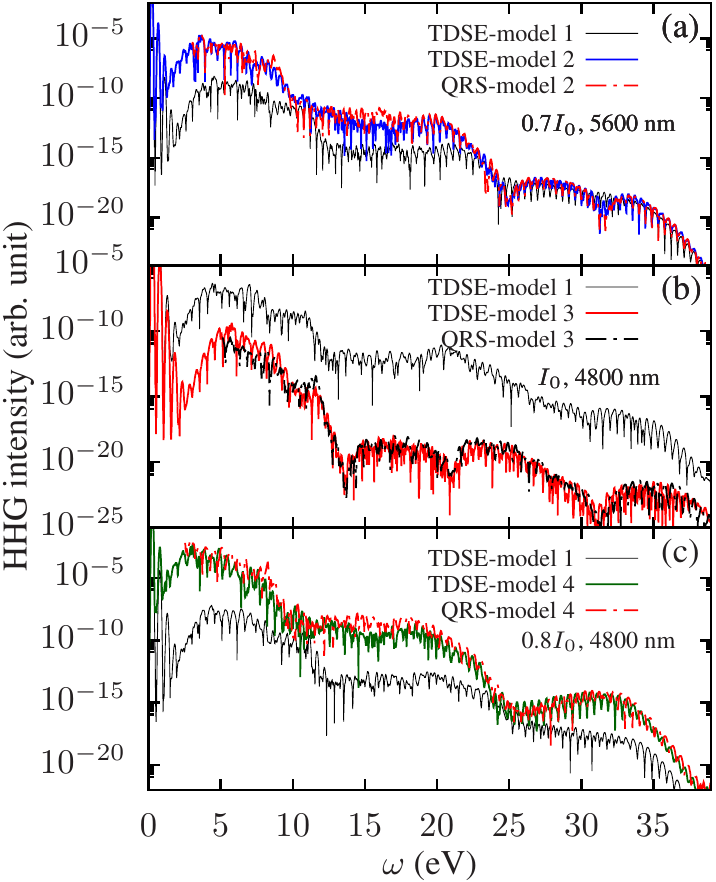}}
	\caption{(a) Comparison of HHG spectra from model 1 and model 2 obtained from the TDSE with the same 10-cycle laser pulse at wavelength of $5.6$ $\mu$m and intensity of $7\times10^{10}$ W/cm$^2$. The QRS result for model 2 is shown as the dashed line. (b) Same as (a) but for model 1 and 3 for the laser pulse at wavelength of $4.8$ $\mu$m and intensity of $10^{11}$ W/cm$^2$. (c) Same as (a) but for model 1 and 4 for the laser pulse at wavelength of $4.8$ $\mu$m and intensity of $8\times10^{10}$ W/cm$^2$.}
	\label{Fig3}
\end{figure}
\begin{figure}[htbp]
	\centering
	\fbox{\includegraphics[width=0.9\linewidth]{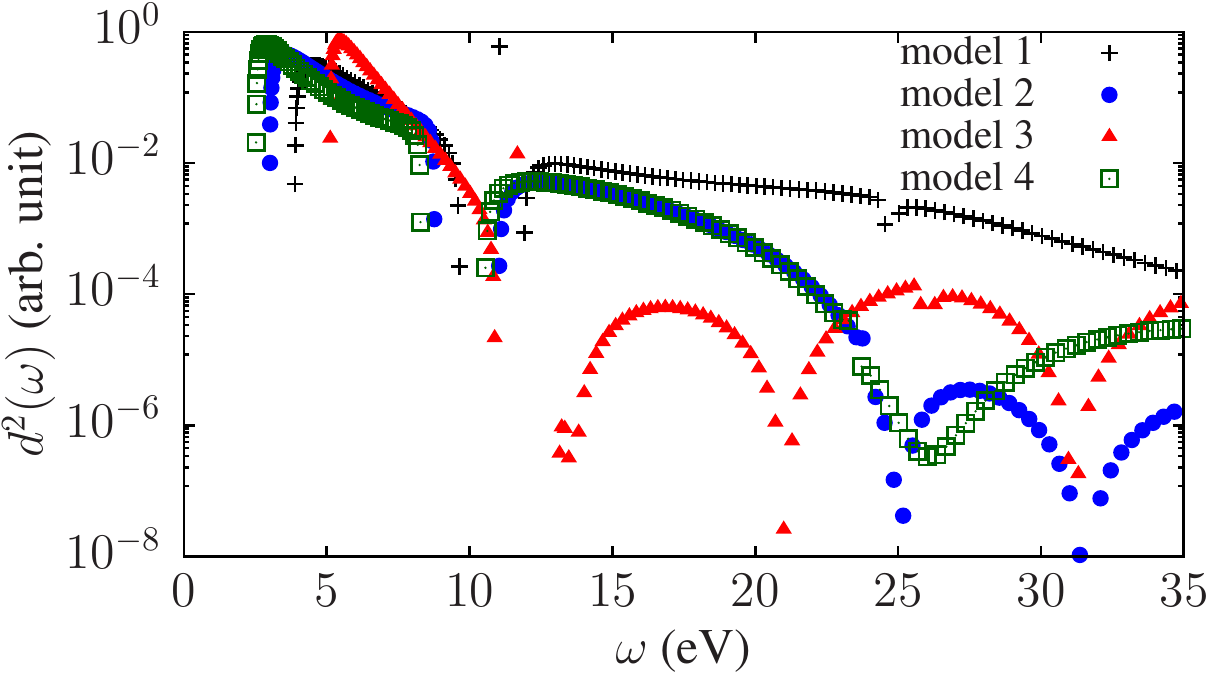}}
	\caption{Modulus squared of transition dipoles from the impurity ground state to the conduction bands versus photon energy for model 1, 2, 3 and 4.}
	\label{Fig4}
\end{figure}

In the following, we will use model 1 as a reference system for comparison with the other models. In Fig.~\ref{Fig3}(a) we compare HHG spectra obtained from the TDSE for model 1 and model 2 under the same 10-cycle laser pulse at wavelength of $5.6$ $\mu$m and intensity of $7\times10^{10}$ W/cm$^2$. There are three distinct energy regions (or plateaus) above the threshold at about 5 eV. By comparing with Fig.~\ref{Fig2}, the first plateau (from 5 eV to about 10 eV) can be associated with the recombination from CB1 back to the impurity ground state. Similarly, the second and third plateau, from 13 eV to 22 eV and from 25 eV to 35 eV, respectively, is associated with the recombination from CB2 and CB3 to the impurity, respectively. For each model, there are significant drops in HHG yields for energy regions between the plateaus. This can be attributed to the exponential decrease of tunneling excitation from a lower conduction band to the next one. Note that the first plateau in model 1 is slightly more extended than that of model 2. This is due to the presence of an impurity level in the energy gap between CB1 and CB2 in model 1.

More importantly, for the harmonics below 20 eV, the HHG yields from model 2 is about two orders of magnitude stronger than that of model 1. This is not entirely surprising, since the energy gap between the highest occupied orbital of the impurity and CB1 in model 2 is about 2 eV smaller (see Fig.~\ref{Fig2}). However, for the energies above 23 eV, the HHG yields from the two models are nearly identical. 

To understand the origin of this behavior, we compare in Fig.~\ref{Fig4} the modulus square of the transition dipoles from the conduction bands to the impurity ground state for the two models, as they are proportional to the photo-excitation and photo-recombination cross sections. For the energies below 20 eV, the cross sections are quite similar. Therefore the difference in the HHG yields from the two models could be attributed mainly to the differences in tunneling excitation (or ``ionization'') from the impurity ground state to CB1, which in turn depends on the energy gaps as discussed above. For the energies above 23 eV, the cross section from model 1 is about two order of magnitude stronger than that of model 2. That somewhat compensates the weaker ``ionization'' in model 1 so that the HHG yields from the two models are nearly identical above 23 eV. Furthermore, a closer look at HHG spectrum from model 2 reveals two weak minima near 25 eV and 31 eV, at the same energies that its cross section has minima. In contrast, model 1 shows no obvious minimum in that energy range in both HHG and cross section.  

Similar comparisons for model 3 and model 4 with model 1 are showed in Fig.~\ref{Fig3}(b) and Fig.~\ref{Fig3}(c), respectively. The laser parameters are given in the figure caption. Here, the locations of the plateaus are quite similar to that of Fig.~\ref{Fig3}(a). More importantly, for all the cases, the structures in HHG spectra follow closely the cross sections, shown in Fig.~\ref{Fig4}. In particular, the pronounced minima in the HHG spectrum for model 3 near 13 eV, 21 eV, and 31 eV are clearly associated with the three minima in the cross sections at the same energies. For model 4, the broad minimum in the HHG spectrum near 26 eV is caused by the respective minimum in the cross section. Again, there is no obvious minimum in the same energy regions in both HHG and cross section for model 1. 

The above analysis indicates a close relationship between HHG and photo-recombination cross sections. This strongly suggests that the QRS \cite{Morishita:prl08,Le:pra09,Lin:jpb10,Le:jpb16} can be extended for impurities in solids. According to the QRS theory, HHG yields can be calculated as a product of an electron wave-packet and the photo-recombination cross section. Since the conduction bands are not affected much by the presence of the impurities, one can assume that the electron wave packets for different impurities are nearly identical (up to overall constants, which account for different excitation probabilities from the impurity ground states to the conduction bands). Therefore, by using the QRS, one can, for example, easily obtain a HHG spectrum for one target, if a HHG spectrum for another target (the reference) under the same laser is known \cite{Morishita:prl08,Le:pra09,Lin:jpb10}. The QRS results are shown in Fig.~\ref{Fig3} as dashed lines. In all the cases, they agree very well with the exact TDSE for the whole energy region above the threshold (near 5 eV). Here, model 1 was used as the reference target. 

\begin{figure}[t]
	\centering
	\fbox{\includegraphics[width=0.9\linewidth]{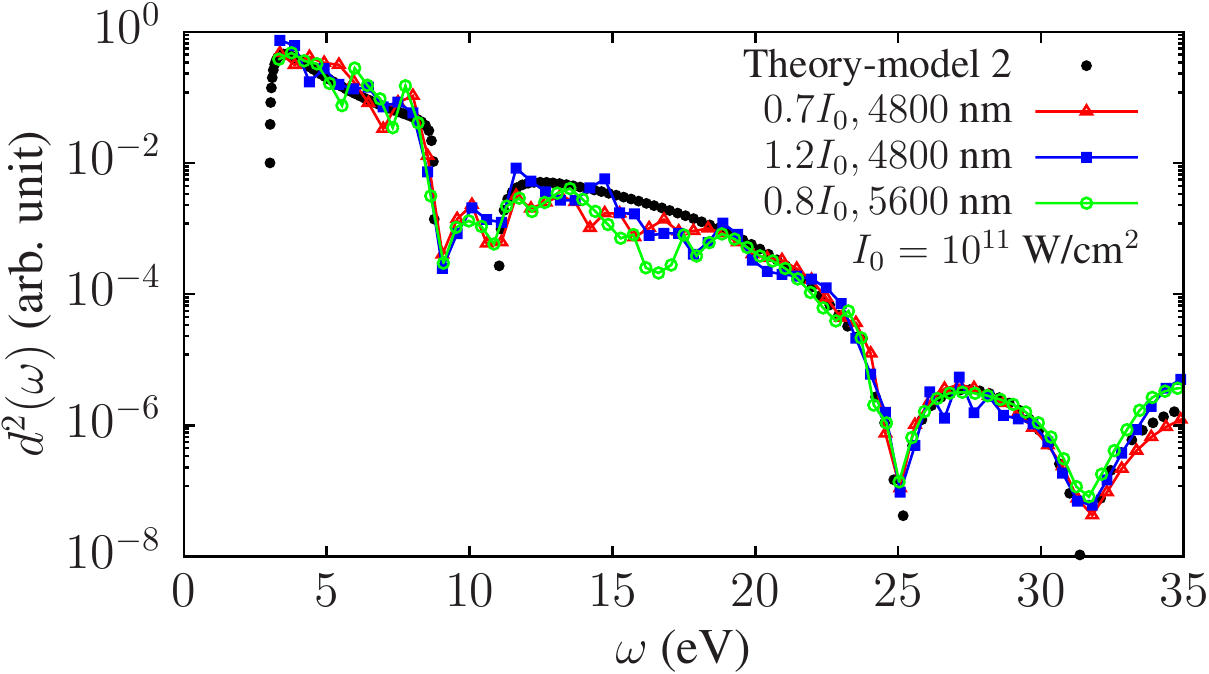}}
	\caption{Retrieved modulus squared of transition dipole from the impurity ground state to the conduction bands versus photon energy for model 2 using HHG with different lasers. The laser parameters are given in the labels. Here, $I_0=10^{11}$ W/cm$^2$. Theoretical result for laser-free transition dipole is also shown as black solid dots.}
	\label{Fig5}
\end{figure}

Using the QRS in the same manner as it is done for gases, one can also extract the photo-recombination cross section for a unknown impurities target, if HHG spectra for the target and a reference target are known. We show the results for model 2 in Fig.~\ref{Fig5}. The retrieved transition dipoles squared, obtained from HHG spectra with different lasers, agree well with the theoretical data (black solid dots) for a broad range of energy from the threshold to 35 eV. In particular, the minima near 25 eV and 32 eV are nicely retrieved. Here, we use model 1 as the reference target. We remark that  photoionization cross sections for atoms and molecules in the gas phase have been retrieved experimentally from HHG measurements by using the same method, see for example, Refs.~\cite{Minemoto:pra08,Shiner:NatPhys11,Higuet:pra11}. Quite recently, a tomographic reconstruction of impurity orbitals using HHG was suggested based on the three-step model \cite{Almalki:prb18}.    

Our microscopic treatment of HHG needs to be complemented by macroscopic propagation simulations for realistic comparisons with experiments. In principle, this can be done by solving coupled TDSE and Maxwell's equations. Progress along this direction was reported recently \cite{Floss:pra18}. 

In conclusions, we have established that HHG process from impurities in doped materials can be expressed as a product of an electron wave packet and transition dipole for the electron in a conduction band back to the impurity ground state. This process therefore resembles very closely the HHG process in atoms and molecules in the gas phase, except that the conduction bands now play the role of the continuum. Based on this approximate factorization, we have extended the quantitative rescattering theory for HHG from impurities in solids. We expect that our theory can serve as a simple starting point to study HHG in solids for realistic systems. Research along this direction could provide detailed information about impurities in solid environment, needed for understanding and controlling various processes in engineered solid structures.        
	
\noindent \textbf{Funding.} V. H. H. is funded by the Program of Fundamental Research of Ministry of Education and Training (Vietnam) under Grant No. B2016.19.10., and by the Vietnam National Foundation for Science and Technology Development (NAFOSTED) under Grant No. 103.01-2017.371. The computing part of this project was performed on the Forge cluster at Missouri University of Science and Technology.    

\noindent \textbf{Acknowledgment.} A.T.L. thanks Lars Bojer Madsen, Chuan Yu, and Kenneth K. Hansen for valuable and insightful discussions. 

\noindent \textbf{Disclosures.} The authors declare no conflicts of interest

\bibliography{MyBib}
\bibliographystyle{apsrev4-1}

\end{document}